\documentclass[a4paper,fleqn]{cas-dc}

\usepackage[numbers,sort&compress]{natbib}

\def\tsc#1{\csdef{#1}{\textsc{\lowercase{#1}}\xspace}}
\tsc{WGM}
\tsc{QE}
\tsc{EP}
\tsc{PMS}
\tsc{BEC}
\tsc{DE}

\begin{document}
\let\WriteBookmarks\relax
\def\floatpagepagefraction{1}
\def\textpagefraction{.001}

\shorttitle{High-Flux Cold Ytterbium Atomic Beam Source Using Two-Dimensional Laser Cooling with Intercombination Transition}

\shortauthors{ T. Hosoya \it et~al.}

\title [mode = title]{High-Flux Cold Ytterbium Atomic Beam Source Using Two-Dimensional Laser Cooling with Intercombination Transition}                      



%
\author[1]{Toshiyuki Hosoya}






\affiliation[1]{organization={Department of Physics, Tokyo Institute of Technology},
    addressline={2-12-1 Ookayama, Meguro-ku}, 
    city={Tokyo},
    postcode={152-8550}, 
    country={Japan}}

\author[2]{Ryotaro Inoue}

\author[2]{Tomoya Sato}


\affiliation[2]{organization={Institute of Innovative Research, Tokyo Institute of Technology},
    city={4259 Nagatsuta-cho, Midori-ku, Yokohama, Kanagawa},
    postcode={226-8503}, 
    country={Japan}}

\author%
[1,2]
{Mikio Kozuma}
\cormark[1]

\cortext[cor1]{Corresponding author. Tel./fax: +81 45 924 5435.\\
Email address: kozuma@qnav.iir.titech.ac.jp (Mikio Kozuma)}

\begin{abstract}
We demonstrate a high-flux and low transverse temperature atomic beam of ytterbium by applying two-dimensional cooling using the ${}^1\mathrm{S}_0\text{-}{}^3\mathrm{P}_1$ intercombination transition to the cold atomic beam produced by the dipolar-allowed ${}^1\mathrm{S}_0\text{-}{}^1\mathrm{P}_1$ transition.
The optimized transverse temperature of $11 \pm 9\,\mathrm{\mu K}$ and an atomic flux of $(7.5 \pm 1.0) \times 10^8\,\mathrm{atoms/s}$ are obtained for the atomic beam whose longitudinal velocity is $30\,\mathrm{m/s}$. 
The transverse temperature is maintained below $23\,\mathrm{\mu K}$, while the flux is above $6.5 \times 10^8\,\mathrm{atoms/s}$ in the longitudinal velocity range of $22 - 30\,\mathrm{m/s}$.
We also discuss the feasibility of further narrowing the transverse momentum width to less than the recoil momentum using the momentum-selective optical transition between the ground state and the long-lived metastable state. The transverse momentum width of our atomic beam, which is narrower than 5.7 times the recoil momentum, is a good starting point for the proposed method. 
\end{abstract}




\maketitle

\section{Introduction}\label{sec:int}

Atom interferometry is a powerful tool for precision measurements~\cite{Zhou2015,Bouchendira2011,Olson2019} that enables various inertial sensing applications, such as gravity accelerometers~\cite{Kasevich1992,Peters1999,Mueller2008,Mazzoni2015}, gravity gradiometers~\cite{Snadden1998,Mcguirk2002,Bertoldi2006,Sorrentino2014,Damico2016,DelAguila2018}, and gyroscopes~\cite{Riehle1991,Gustavson1997,Gustavson2000,Canuel2006,Gauguet2009,Savoie2018,Xue2015}.
While the pioneering experiments mainly deal with alkali atoms, there has been progress in the construction of a next-generation atom-interferometry-based inertial sensor using two-electron atomic species~\cite{Mazzoni2015,DelAguila2018} such as strontium (Sr) and ytterbium (Yb).
The interferometer, which exploits their spin-singlet ground states that are insensitive to magnetic fields, can potentially prevent the sensing stability from deteriorating due to environmental magnetic fields.
In addition, the use of higher-order Bragg diffraction~\cite{Mazzoni2015} will further improve the sensitivity because its large momentum transfer enables us to increase the area of the interferometer.

More recently, atom interferometry with a laser-cooled slow atomic beam~\cite{Kwolek2020} for continuous or dead-time-free data acquisition has attracted attention for use in practical applications. For interferometry with the Bragg diffraction technique, the transverse momentum width of the atomic beam should be less than $0.25\hbar k$~\cite{Szigeti2012} to maintain the efficiency of Bragg diffraction, where $\hbar k$ is the recoil momentum of the Bragg beam.
Such an atomic beam with an extremely narrowed transverse momentum width has been achieved by applying the momentum selection method to the thermal atomic beam~\cite{Giltner1995}; a pair of slits filter out the unwanted momentum components at the expense of its beam flux.
To make the atomic beam interferometer with Bragg diffraction practical for inertial sensing, the essential requirements that have to be satisfied by the beam include a narrow transverse momentum width and a high beam flux. Here, we study the two-dimensional laser cooling of a ${}^{174}$Yb atomic beam using the ${}^1\mathrm{S}_0\text{-}{}^3\mathrm{P}_1$ intercombination transition, which can reduce the transverse momentum width to $2.3 \hbar k_{399}$~\cite{Gomes2017} in principle, where $\hbar k_{399}$ is the recoil momentum of the dipolar-allowed ${}^1\mathrm{S}_0\text{-}{}^1\mathrm{P}_1$ transition of Yb.
In this work, we successfully suppressed the transverse momentum width of $(4.0\pm 1.6)\hbar k_{399}$ with an atomic flux of $(7.5 \pm 1.0) \times 10^8\,\mathrm{atoms/s}$ for an atomic beam with a longitudinal mean velocity of $30\,\mathrm{m/s}$. 
The transverse momentum width is maintained below $5.7\hbar k_{399}$ while maintaining an atomic flux above $6.5 \times 10^8\,\mathrm{atoms/s}$ in the mean longitudinal velocity range of $22-30\,\mathrm{m/s}$.
Additionally, we discuss the feasibility of further narrowing the momentum width by using one of the long-lived metastable states of Yb.

\section{Slow and continuous atomic beam generated by the dipolar-allowed transition}\label{sec:399}

\begin{figure}[tbp]
    \begin{center}
        \includegraphics[width=0.9\linewidth]{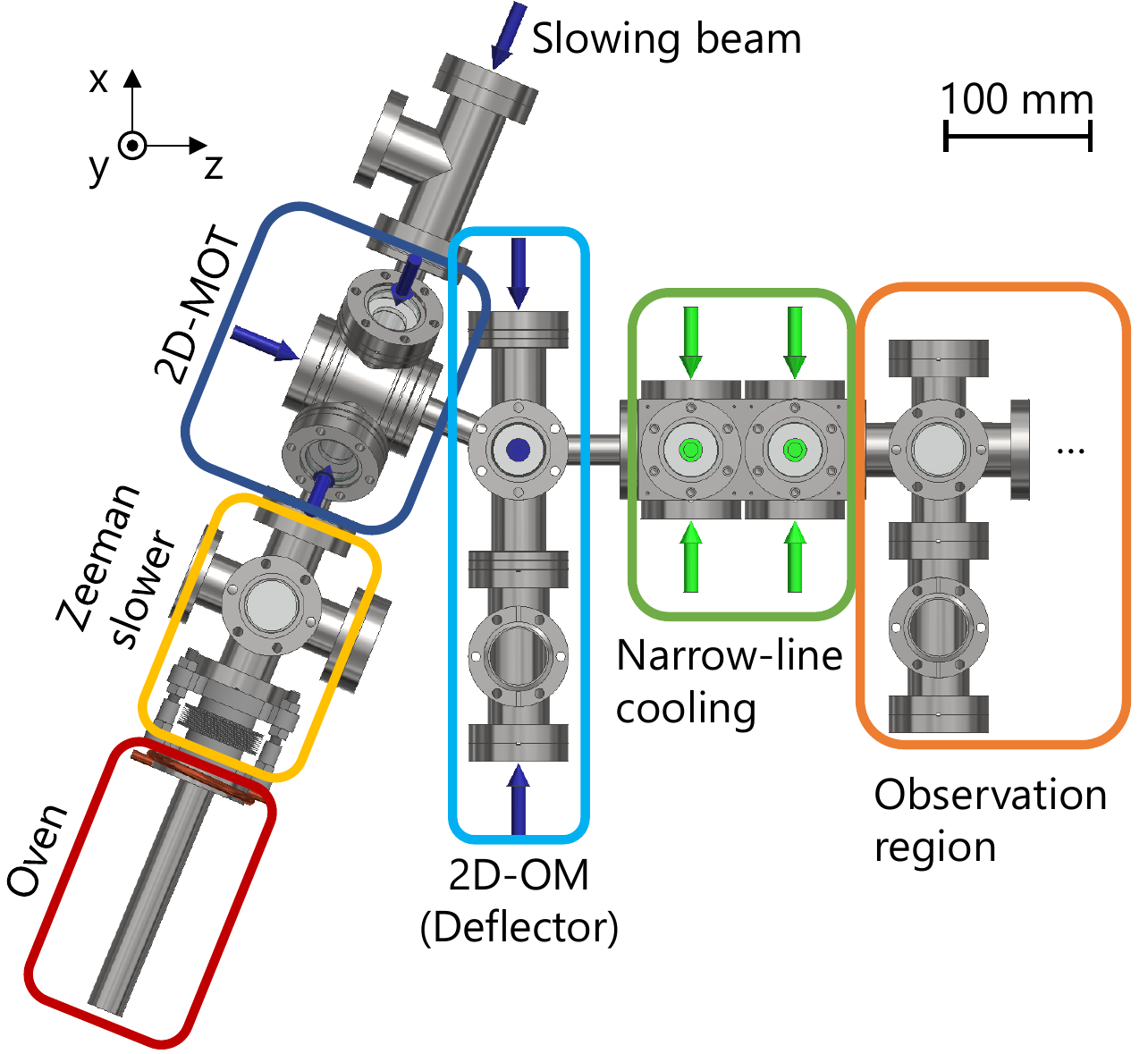}
        \caption{Schematic of the experimental setup. The atoms emitted from the effusive oven are decelerated by the Zeeman slower and are captured by the 2D-MOT. A weak laser beam expels trapped atoms from the 2D-MOT and generates a slow atomic beam. This slow atomic beam is subsequently deflected via off-axis 2D-OM; in particular, it is radially cooled by two 2D-OMs operating on the ${}^1\mathrm{S}_0\text{-}{}^3\mathrm{P}_1$ transition (beams in green).}
        \label{fig:setup}
    \end{center}
\end{figure}

We generate a slow atomic beam with a transverse temperature of a few millikelvin using the dipolar-allowed ${}^1\mathrm{S}_0\text{-}{}^1\mathrm{P}_1$ transition.
Fig.~\ref{fig:setup} shows a schematic of our experimental setup.
A thermal atomic beam from the effusive oven at $650\,\mathrm{K}$ is decelerated by a Zeeman slower (ZS).
The decelerated atoms are cooled and captured with a two-dimensional magneto-optical trap (2D-MOT).
Then, the trapped atoms are expelled by a weak laser beam (pushing beam).
Four permanent magnets produce a magnetic field for the ZS and the 2D-MOT, similar to the 2D-MOT system for the cold Sr atom source~\cite{Nosske2017}.
The maximum strength of the magnetic field for our ZS is approximately $21\,\mathrm{mT}$, and the magnetic field gradient for the 2D-MOT is $400\,\mathrm{mT/m}$.
Each laser beam for this stage is red-detuned from the ${}^1\mathrm{S}_0\text{-}{}^1\mathrm{P}_1$ transition at a wavelength of $399\,\mathrm{nm}$ (natural linewidth of $\Gamma_{399} / 2 \pi=29\,\mathrm{MHz}$ and saturation intensity of $I_{s,399} = 57\,\mathrm{mW/cm^2}$). 
The detuning of the ZS beam is set to $-10\Gamma_{399}$ according to the maximum Zeeman shift of $-9.8\Gamma_{399}$ in our ZS.
The cooling light for our 2D-MOT is also set to a peak intensity of $1.5 I_{s,399}$ and a detuning of $-1.0\Gamma_{399}$ to maximize the expelling flux of the slow atomic beam.
The diameters of the ZS and 2D-MOT beams are designed to match the shape of the thermal atomic beam from our effusive oven.
We set a $1/e^2$ diameter of $11\,\mathrm{mm}$ for the ZS beam at the 2D-MOT region (focused on the oven outlet at $300\,\mathrm{mm}$ away) and of $7.8\,\mathrm{mm}$ ($\sim 11\, \mathrm{mm}/ \sqrt{2} $) for the 2D-MOT beams.

The subsequent two-dimensional optical molasses (2D-OM) deflects the incoming slow atomic beam by an angle of $22.5^\circ$ and separates the outgoing atomic beam from the pushing light beam.
We perform the 2D-OM using two sets of retroreflecting light beams propagating along the horizontal (x in Fig.~\ref{fig:setup}) and vertical (y in Fig.~\ref{fig:setup}) axes.
The horizontal and vertical beams have peak intensities of $2.3 I_{s,399}$ and $0.23 I_{s,399}$, respectively.
Each beam has a $1/e^{2}$ diameter of $9.7\,\mathrm{mm}$ and is detuned $-1.0\Gamma_{399}$ from the resonance frequency.
While the detuning of $-\Gamma_{399}/2$ brings the minimum transverse temperature of the output atomic beam in principle, the detuning causes a considerable loss of the beam flux in our experiment due to the oblique incidence of the atoms.
For example, an obliquely incoming Yb atom with a longitudinal velocity of $30\,\mathrm{m/s}$ gives the Doppler shift of $\sim\Gamma_{399}$ taking into account the tilting angle of $22.5^\circ$.
We test the deflection of the atomic beam with $\mathrm{lin}\perp\mathrm{lin}$ (2D-OM configuration~\cite{Dammalapati2009}) and $\sigma^+ ~ \text{-} ~ \sigma^-$ (2D-MOT configuration~\cite{Yang2015}) light beam polarizations with a magnetic field gradient of $140\,\mathrm{mT/m}$. 
Here, we adopt the 2D-OM configuration since we find that the pointing of the atomic beam from the 2D-MOT-type deflector varies depending on the longitudinal velocity of the incoming atoms. We hypothesize that the dependency is due to the insufficient length of our deflector region.
In this experimental study, we utilized an atomic beam with a mean longitudinal velocity of $30\,\mathrm{m/s}$ for the overall parameter optimizations.

The narrow linewidth of a ${}^1\mathrm{S}_0\text{-}{}^3\mathrm{P}_1$ intercombination transition at a wavelength of $556\,\mathrm{nm}$ (the natural linewidth $\Gamma_{556}/2\pi = 182\,\mathrm{kHz}$, the saturation intensity $I_{s,556}=0.14\,\mathrm{mW/cm^2}$) enables us to probe the transverse velocity distribution of the pre-cooled atomic beam by evaluating the Doppler broadened spectrum.
The spectroscopy gives the transverse velocity width $\sigma_v^{(0)}=0.28(1)\,\mathrm{m/s}$, which corresponds to the transverse momentum width of $(48 \pm 2) \hbar k_{399}$ and the transverse temperature of $1.6(1)\,\mathrm{mK}$.
Here, we use the half-width at $1/\sqrt{e}$ of the peak flux density to characterize the transverse width.
The $556\,\mathrm{nm}$ light is generated by a frequency doubling of a fiber laser output operated at a wavelength of $1112\,\mathrm{nm}$ and a linewidth of less than $20\,\mathrm{kHz}$, which is also utilized in the following narrow-line cooling experiment.

\section{Transverse cooling with an intercombination transition}\label{sec:556}

\begin{figure*}[htbp]
    \begin{center}
        \includegraphics[width=0.8\linewidth]{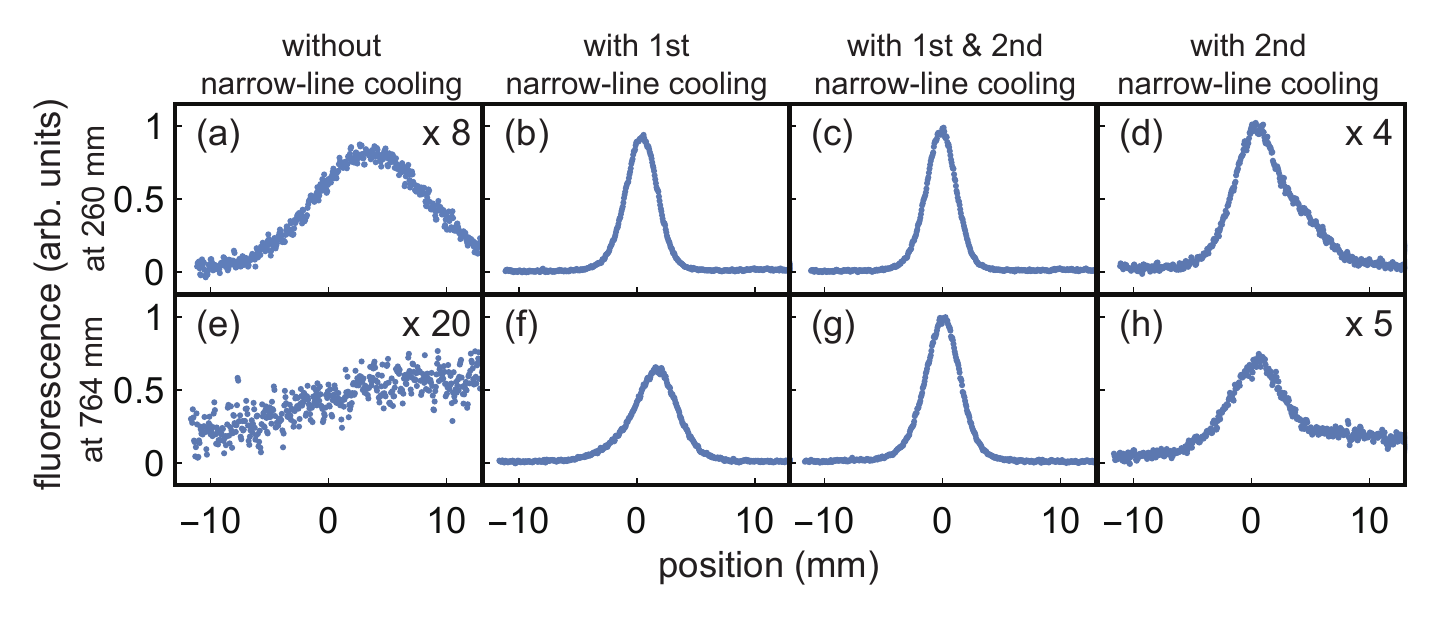}
        \caption{Effect of narrow-line cooling on the spatial distribution of the atomic beam. The distributions are measured based on fluorescence imaging of the ${}^1\mathrm{S}_0\text{-}{}^1\mathrm{P}_1$ transition, with a CMOS camera placed (a) - (d) $260\,\mathrm{mm}$ and (e) - (h) $764\,\mathrm{mm}$ downstream of the second narrow-line cooling region. The results are measured (a), (e) without narrow-line cooling, (b), (f) with the first narrow-line cooling laser alone, (c), (g) with both the first and second narrow-line cooling lasers, and (d), (h) with the second narrow-line cooling laser alone.}
        \label{fig:dist}
    \end{center}
\end{figure*}

The transverse temperature of the atomic beam is further cooled by using the ${}^1\mathrm{S}_0\text{-}{}^3\mathrm{P}_1$ transition.
The minimum transverse temperature of $3.7\,\mathrm{\mu K}$~\cite{Gomes2017} (the corresponding velocity width is $0.013\,\mathrm{m/s}$) is expected for two-dimensional cooling when the detuning is $-\Gamma_{556} /2$ and the intensity is much smaller than $I_{s,556}$.
To evaluate the resultant velocity distribution, we measure the ballistic expansion of the atomic beam under various conditions, as shown in Fig.~\ref{fig:dist}.
The upper panels (a)-(d) and the lower panels (e)-(h) are the spatial distributions obtained by fluorescence imaging with a camera placed at $260\,\mathrm{mm}$ and $764\,\mathrm{mm}$ downstream of the second narrow-line cooling region, respectively.
The left-most panels (a) and (e) are the results without narrow-line cooling, which show broad distributions due to a broad transverse velocity width of $\sigma_v^{(0)}=0.28(1)\,\mathrm{m/s}$.

In conventional experiments, the cooling parameters such as laser intensity, detuning, and magnetic field gradient are temporally varied during narrow-line cooling to achieve a broader capture range and a lower temperature simultaneously~\cite{Doerscher2013,Kemp2016}.
As shown in Fig.~\ref{fig:setup}, our narrow-line cooling stage consists of two regions to enable us to change the cooling parameters during the flight of the atoms.
The first and second narrow-line cooling beams have $1/e^2$ diameters of $16\,\mathrm{mm}$ and $10\,\mathrm{mm}$, respectively.
Both light beams are sufficiently larger than the spatial spreading of the atomic beam at these cooling regions, which is characterized by a $1/\sqrt{e}$ radius of $1.2(1)\,\mathrm{mm}$.
Note that we choose the $\mathrm{lin}\perp\mathrm{lin}$ polarization configuration (2D-OM) for the sake of simplicity.\footnote{The $\sigma^+ ~ \text{-} ~ \sigma^-$ configuration with a magnetic field gradient (2D-MOT) might be an alternative to increase the velocity capture range and compress the outgoing atomic beam.}\label{fot:sig}

\begin{figure}[htbp]
    \begin{center}
        \includegraphics[width=0.9\linewidth]{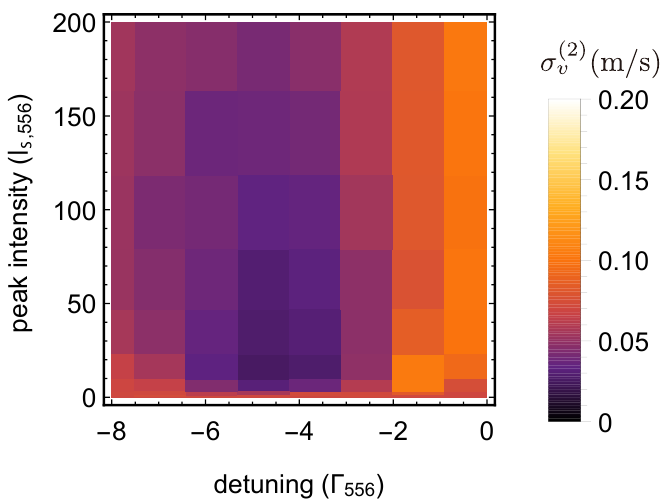}
        \caption{The transverse velocity width of outgoing atoms $\sigma _\nu^{(2)}$ as a function of the peak intensity and the detuning of the cooling beams in the second narrow-line cooling region.}
        \label{fig:dens}
    \end{center}
\end{figure}

The Gaussian intensity distribution of the cooling beams also broadens the velocity capture range while maintaining the low transverse temperature since atoms experience the laser intensity variation by simply crossing the cooling laser.
Numerical simulations of the atomic motion predict that we can achieve a velocity capture range of $0.86\,\mathrm{m/s}>\sigma_v^{(0)}$ and an outgoing transverse velocity width of $0.055\,\mathrm{m/s}$ by using the cooling light, which has a peak intensity of $300I_{s,556}$ and a detuning of $-7.9\Gamma_{556}$.
The experimental results with the above cooling parameters ( $-7.9\Gamma_{556},300I_{s,556}$) are shown in Fig.~\ref{fig:dist}(b) and (f).
The extracted transverse velocity width $\sigma_v^{(1)}=0.063(5)\,\mathrm{m/s}$ has no discrepancy with the numerically obtained value.
For the second narrow-line cooling, the theoretical model predicts that the detuning of $-\Gamma_{556}/2$ with the peak intensity range of $10I_{s,556}-30I_{s,556}$ achieves the (near-)Doppler-limited transverse velocity width of $0.013\,\mathrm{m/s}$ with the velocity capture range of $0.3\,\mathrm{m/s}>\sigma_v^{(1)}$.
Fig.~\ref{fig:dens} summarizes the experimentally extracted transverse velocity width $\sigma_v^{(2)}$ after the second narrow-line cooling as a function of the peak intensity and the detuning of the second narrow-line cooling beams.
We find that the set of cooling parameters $(-3.7\Gamma_{556},\ 14I_{s,556})$ minimizes $\sigma_v^{(2)}$ to $0.023(9)\,\mathrm{m/s}$.
The discrepancy in the optimal detuning with the numerically obtained value can be explained by assuming the misalignment of counterpropagating cooling beams, which shifts the optimal detuning due to the atomic longitudinal velocity.
Under the optimal cooling condition $(-3.7\Gamma_{556},\ 14I_{s,556})$, we obtain the transverse velocity width $\sigma _v^{(2)}=0.023(9)\,\mathrm{m/s}$ corresponding to a transverse temperature of $11 \pm 9\,\mathrm{\mu K}$ or a transverse momentum width of $(4.0 \pm 1.6) \hbar k_{399}$ with an atomic flux of $(7.5 \pm 1.0) \times 10^8\,\mathrm{atoms/s}$.
The near-indistinguishable profiles shown in Fig.~\ref{fig:dist}(c) and (g) indicate successful narrow-line cooling.
On the other hand, under the same cooling conditions, the blocking of the first narrow-line cooling causes a drastic deterioration of the atomic beam quality, as shown in Fig.~\ref{fig:dist}(d) and (h).
The bimodal characteristics of the distributions suggest that most atoms were simply passing through, exceeding the velocity capture range of the second narrow-line cooling.
The results indicate that our cascaded setup for narrow-line cooling is essential to achieve both a wider capture range and lower transverse temperature in our system.

\section{Tunability and stability of longitudinal velocity}\label{sec:vel}

\begin{figure*}[htbp]
    \begin{center}
        \includegraphics[width=0.9\linewidth]{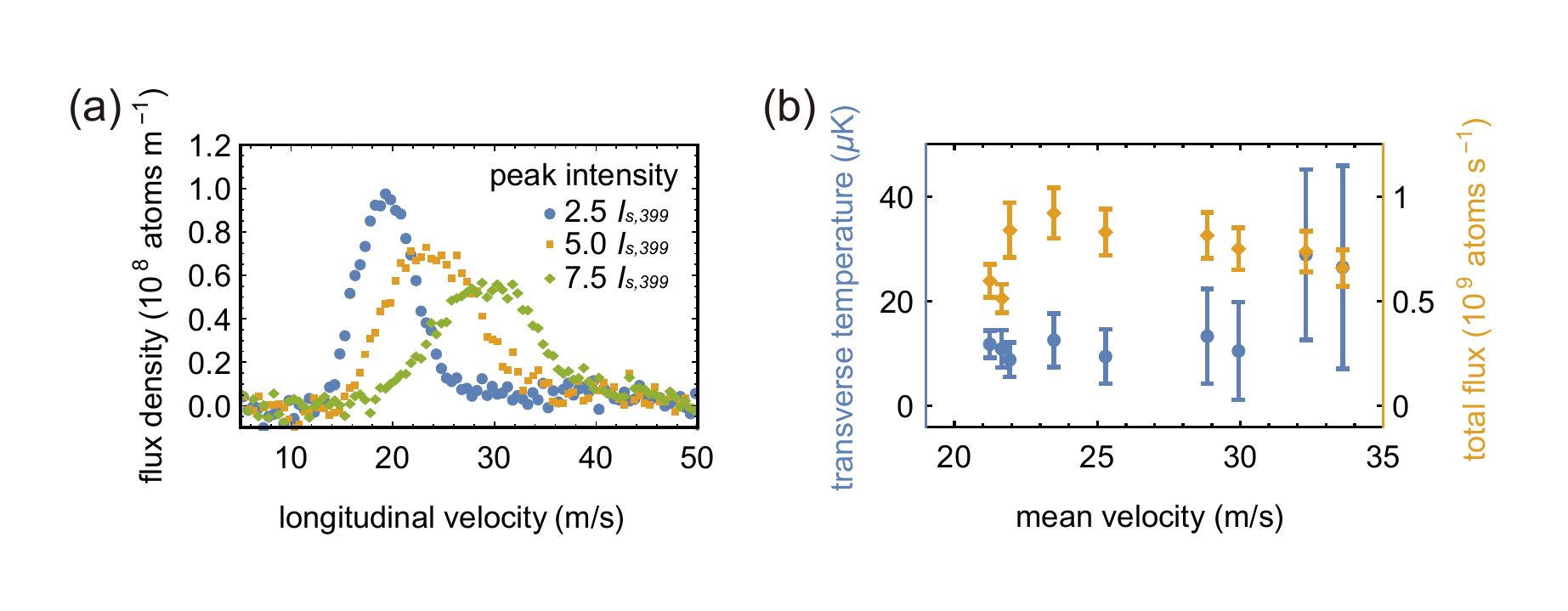}
        \caption{(a) Longitudinal velocity distribution of the atomic beam for the pushing beam with a detuning of -$1.7 \Gamma_{399}$ and peak intensity of $2.5I_{s,399}$ (blue circles), $5.0I_{s,399}$(orange squares), and $7.5I_{s,399}$(green rhombus). (b) Dependence of the transverse temperature (blue circles) and total flux (orange rhombus) on the longitudinal mean velocity with an oven temperature of $650\,\mathrm{K}$ and a ZS laser power of $80\,\mathrm{mW}$.}
        \label{fig:long}
    \end{center}
\end{figure*}

\begin{figure}[htbp]
    \begin{center}
        \includegraphics[width=0.8\linewidth]{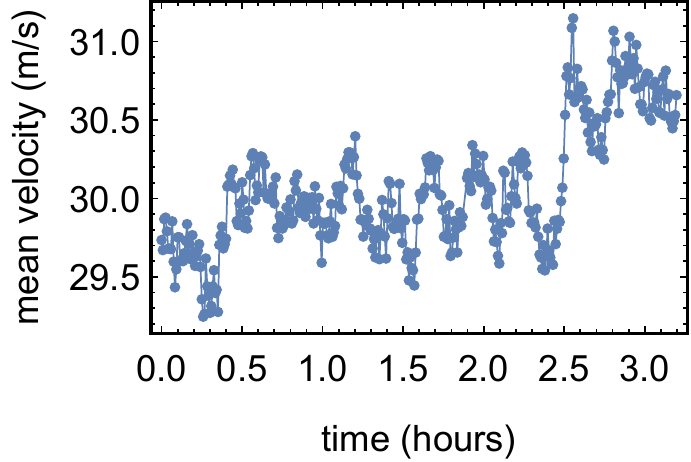}
        \caption{Mean longitudinal velocity as a function of time for the pushing beam with a peak intensity of $7.5I_{s,399}$ and detuning of $-1.7\Gamma_{399}$.}
        \label{fig:mean}
    \end{center}
\end{figure}

Using the intensity of the pushing beam as a tuning parameter, Fig.~\ref{fig:long} demonstrates the tunability of the longitudinal velocity in our setup.
We obtain the longitudinal velocity distribution from the time-of-flight (TOF) measurements described below.
A plug beam resonant on the ${}^1\mathrm{S}_0\text{-}{}^1\mathrm{P}_1$ transition, which blocks the atomic beam, is positioned at $260\,\mathrm{mm}$ downstream of the second narrow-line cooling.
We generate the pulsed atomic beam by turning off the plug beam for $0.5\,\mathrm{ms}$.
The corresponding pulsed fluorescence is measured at $764\,\mathrm{mm}$ downstream of the second narrow-line cooling region.
We then extract the longitudinal velocity distribution from the obtained fluorescence signals; Fig.~\ref{fig:long}(a) shows the typical results.
The mean of the longitudinal velocity distribution changes from $22\,\mathrm{m/s}$ to $30\,\mathrm{m/s}$ as the intensity of the pushing beam increases from $2.5I_{s,399}$ to $7.5I_{s,399}$, in which the full-width at half-maximum is approximately $1/3$ of the mean velocity.
The time series chart shown in Fig.~\ref{fig:mean} also represents the typical stability of the mean velocity, which is measured every $30\,\mathrm{s}$ for $3.2\,\mathrm{h}$ using the TOF method.
To apply the atomic beam to the atom-interferometer-based inertial sensor, the dynamic range~\cite{Kwolek2020} and the scale-factor stability~\cite{Gustavson2000} worsen with an increasing longitudinal velocity width and mean-velocity fluctuation.
Our next challenge will include improving these beam parameters by applying a 3D-OM~\cite{Kwolek2020} approach for our 2D-OM-deflector and narrow-line transverse cooling.

We also measure the dependence of the total flux and the transverse temperature on the longitudinal velocity. The data in Fig.~\ref{fig:long}(b) show that the transverse temperature and the atomic flux are virtually unchanged for the pushing beam intensity corresponding to the mean longitudinal velocities within the $22 - 30\,\mathrm{m/s}$ range.
For the transverse temperature, we obtain $\sim10\,\mathrm{\mu K}$ with a velocity below $30\,\mathrm{m/s}$.
Since we optimized the cooling parameters with a longitudinal velocity of $30\,\mathrm{m/s}$, slower atoms can be cooled efficiently due to the longer cooling time.
Taking into account the errors, our narrow-line cooling with the parameters mentioned above maintains the transverse temperature below $23\,\mathrm{\mu K}$, which corresponds to a momentum width narrower than $5.7 \hbar k_{399}$, and the flux above $6.5 \times 10^8\,\mathrm{atoms/s}$ within the mean velocity range of $22 - 30\,\mathrm{m/s}$.

\begin{figure*}[htbp]
    \begin{center}
        \includegraphics[width=0.8\linewidth]{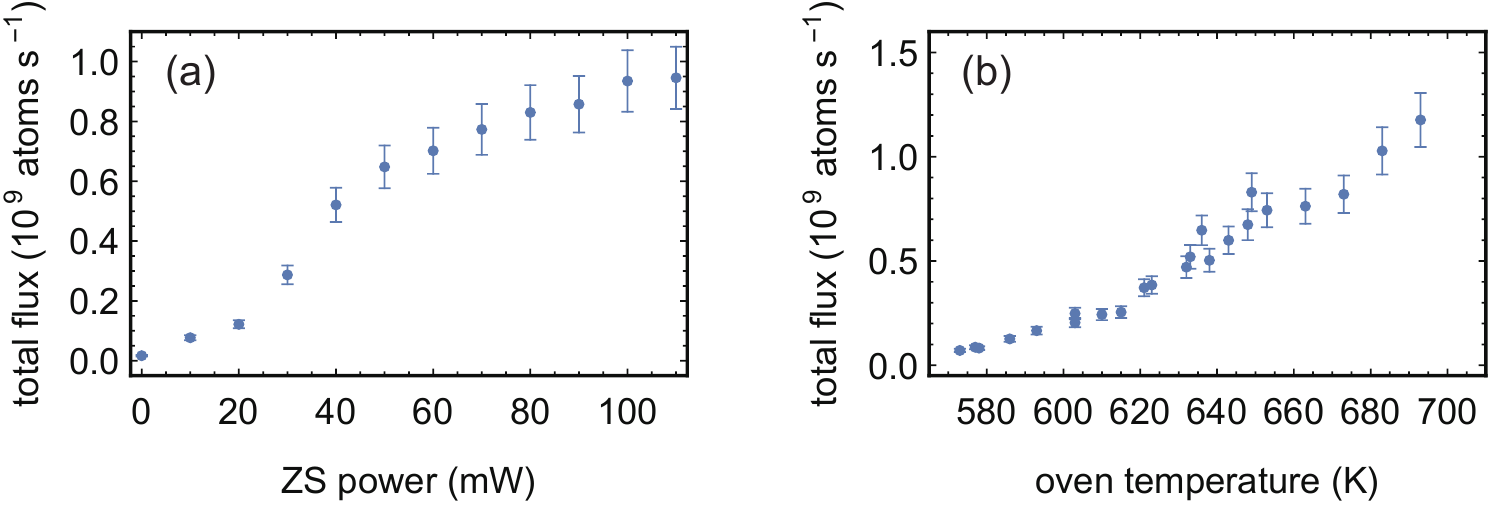}
        \caption{Atomic flux as a function of (a) power of our ZS beam and (b) oven temperature for the pushing beam with peak intensity of $5.0 I_{s,399}$ and detuning of $-1.7 \Gamma_{399}$.}
        \label{fig:flux}
    \end{center}
\end{figure*}

While our research interest is in the narrow-line transverse cooling for the Yb atomic beam described above, an available flux of the atomic beam is also essential for practical applications.
We measure the dependence of the total flux of the narrow-line cooled atomic beam on the power of our ZS beam and the oven temperature (see Fig.~\ref{fig:flux}). 
The total flux is estimated by integrating the flux density obtained by TOF measurements, where the pushing laser intensity is set to $5.0I_{s,399}$ during the evaluation (corresponding mean longitudinal velocity is $25\,\mathrm{m/s}$, see Fig.~\ref{fig:long}(a) for the typical distribution).
In our setup, the total flux shows near-saturation with the ZS laser power of $80\,\mathrm{mW}$, whereas the flux increases monotonically for the oven temperature in the range of $573 - 693\,\mathrm{K}$ with the ZS laser power of $80\,\mathrm{mW}$.
The maximum flux of $(1.2 \pm 0.2) \times 10^9\,\mathrm{atoms/s}$ is obtained at $693\,\mathrm{K}$.

\section{Momentum filtering using an ultra-narrow optical transition}\label{sec:filt}

In higher-order Bragg interferometers, the transverse momentum width of the cold atomic beam must be less than $0.25 \hbar k$~\cite{Szigeti2012}.
One can select the atoms within the narrow momentum width using simple mechanical slits~\cite{Giltner1995} since transverse momentum is correlated with the position of atoms after their long flight.
However, several meters of propagation are required to filter the atoms within the momentum width narrower than the recoil momentum to avoid excessive flux loss.
This issue can be solved by correlating the transverse momentum of atoms with other degrees of freedom.

We propose a scheme to satisfy this requirement using an ultra-narrow optical transition of Yb atoms.
It is supposed that a Yb atomic beam orthogonally crosses the laser light resonant on the ultra-narrow optical transition between the ground state and metastable state, such as ${}^3\mathrm{P}_0$ and ${}^3\mathrm{P}_2$.
When the transit time is much shorter than the lifetime of the metastable state, the transit time determines the spectral linewidth correlated with the transverse momentum width of atoms excited to the metastable state.
In other words, atoms are excited in a momentum-selective manner.
Assuming that the light beam has a Gaussian intensity distribution with a beam waist of $w_l$ along the longitudinal direction of the atomic beam, the resultant momentum distribution of the excited atoms can be expressed by a Gaussian whose standard deviation is $\sigma_p = M v_l / (k_{p} w_l)$~\cite{Demtroeder1981}, where $M$ is the mass of the atom, $v_l$ is the atomic longitudinal velocity, and $k_{p}$ is the angular wavenumber of the light for the momentum selective excitation.
One can obtain an atomic beam with a narrow momentum width by blasting the atoms remaining in the ground state using resonant light on the ${}^1\mathrm{S}_0\text{-}{}^1\mathrm{P}_1$ transition and subsequently returning the atoms to the ground state in the same momentum-selective manner again.
The resultant transverse momentum width $\sigma_p^{\text{(out)}}$ after the overall process can be expressed as $\sigma_p^{\text{(out)}}=\sigma_p/\sqrt{2+(\sigma_p/\sigma_p^{\text{(in)}})^2}$, where $\sigma_p^{\text{(in)}}$ is the transverse momentum width of the incoming atomic beam.
The excessive losses through the filtering process can be minimized by tuning the peak intensity of the light beam so that the integral of the corresponding Rabi frequency with time is equal to $\pi$, i.e., the light beam works as a “$\pi$-pulse”.
By taking into account the inherent loss due to momentum filter, the outgoing fraction of atoms is given by $\sigma_p^{\text{(out)}}/\sigma_{p}^{\text{(in)}}$.
For example, if we intend to distill a momentum-width $\sigma_p^{\text{(out)}} = 0.25 \hbar k_{399}$ component from the narrow-line cooled $^{174}$Yb atomic beam demonstrated in this work, i.e., $\sigma_p^{\text{(in)}} < 5.7 \hbar k_{399}$ and $v_l=30\,\mathrm{m/s}$, the corresponding beam waist $w_l$ with the $^1$S$_0$–$^3$P$_2$ transition at the wavelength of $507\,\mathrm{nm}$ is evaluated as $1.2\,\mathrm{mm}$, and the maximum obtainable fraction of outgoing atoms is estimated to be greater than $4.4\,\mathrm{\%}$; the size of the atomic beam should be much smaller than that of the laser light to achieve the fraction of the outgoing atoms described here.\footnote{The spatial distribution of the atomic beam causes additional losses. For example, the current atomic beam after the narrow-line cooling stage, which has a $1/\sqrt{e}$ radius of $1.2\,\mathrm{mm}$, decreases the expected remaining fraction to $1.5\,\%$ due to the vertical size of the atomic beam.}\label{fot:size}
The required laser power is approximately $10\,\mathrm{W}$\footnote{
The Rabi frequency $\Omega$ of the ${}^1\mathrm{S}_0\text{-}{}^3\mathrm{P}_2$ M2 transition for bosonic isotopes of Yb is given as
$$
   \Omega/2\pi /\sqrt{I}= 2.04\ {\mathrm{Hz}}/\sqrt{\mathrm{W/m^2}}
$$
where $I$ is the intensity of the excitation laser light~\cite{Yamaguchi2008}.
} \label{fot:omega}; such a high-power laser field can be prepared using the intracavity field of a power-built-up cavity~\cite{Antypas2019}.

\section{Conclusion}\label{sec:conc}
We have demonstrated an ytterbium slow atomic beam with an ultra-low transverse temperature.
The beam was produced using three stages: a 2D-MOT, a deflector, and narrow-line cooling.
The narrow-line cooling stage is divided into two regions to achieve a broader capture range and a lower transverse temperature.
The transverse momentum width narrower than $5.7 \hbar k_{399}$ is maintained with a high atomic flux in the longitudinal velocity range of $22 - 30\,\mathrm{m/s}$.
This type of slow atomic beam with a narrow transverse momentum width paves the way for achieveing a continuous Bragg interferometer for inertial sensors.
Our low-divergence atomic beam will also improve the spatial resolution of lithography~\cite{Ohmukai2003}.
Applying additional momentum filtering using a metastable state enables us to obtain the ultra-narrow momentum width of $0.25 \hbar k_{399}$ that is required for a Bragg interferometer.

\section*{Acknowledgement}

This work was supported by JST, Japan Grant Numbers JPMJMI17A3 and JPMJPF2015. T.H. acknowledges partial support from the Japan Society for the Promotion of Science.

\bibliographystyle{elsarticle-num}

\bibliography{refList}

\end{document}